\documentclass[aps,prl,reprint,superscriptaddress,doi=false,isbn=false,url=false,title=true,longbibliography,noeprint]{revtex4-2}
\usepackage{hyperref}
\hypersetup{colorlinks=true, citecolor=blue, urlcolor=blue, linkcolor=blue}
\usepackage{amsmath}
\usepackage{amssymb}
\usepackage{float}
\usepackage{graphicx}
\usepackage{xcolor}
\usepackage{listings}

\newcommand{\chis}{\relax\ifmmode \chi^2/\mathrm{d.o.f.}/\mathrm{d.o.f.}\else $\chi^2/\mathrm{d.o.f.}$ \fi}

\newcommand{\Eth}{\relax\ifmmode E_\mathrm{th}\else $E_\mathrm{th}$\fi}

\begin{document}
\title{Efficient predecision scheme for Metropolis Monte Carlo simulation of long-range interacting lattice systems}

\affiliation{Institut f\"ur Theoretische Physik, Universit\"at Leipzig, IPF 231101, 04081 Leipzig, Germany}
\author{Fabio M\"uller}
\email{fabio.mueller@itp.uni-leipzig.de}
\affiliation{Institut f\"ur Theoretische Physik, Universit\"at Leipzig, IPF 231101, 04081 Leipzig, Germany}
\author{Wolfhard Janke}
\email{wolfhard.janke@itp.uni-leipzig.de}
\affiliation{Institut f\"ur Theoretische Physik, Universit\"at Leipzig, IPF 231101, 04081 Leipzig, Germany}
\date{\today}

\graphicspath{{figures/}}

\begin{abstract}
  We propose a fast and general predecision scheme for Metropolis Monte Carlo simulation of $d$-dimensional long-range interacting lattice models.
For potentials of the form $V(r)=r^{-d-\sigma}$, this reduces the computational complexity from $O\left(N^2\right)$ to $O\left(N^{2-\sigma/d}\right)$ for $\sigma < d$ and to $O\left(N \right)$ for $\sigma > d$, respectively.
The algorithm is implemented and tested for several $\mathrm{O}(n)$ spin models ranging from the Ising over the XY to the Edwards-Anderson spin-glass model.
With the same random number sequence it produces exactly the same Markov chain as a simulation with explicit summation of all terms in the Hamiltonian.
Due to its generality, its simplicity, and its reduced computational complexity it has the potential to find broad application and thus lead to a deeper understanding of the role of long-range interactions in the physics of lattice models, especially in nonequilibrium settings.
\end{abstract}

\maketitle

The Metropolis Monte Carlo algorithm~\cite{Metropolis1953} can be seen as the workhorse for the investigation of lattice systems, allowing the study of both equilibrium properties as well as nonequilibrium processes.
One broad class of models are vector spin ($\mathrm{O}(n)$) models which in various definitions capture diverse physical behavior ranging from first- and second-order phase transitions to infinite-order BKT-type transitions or even glassy behavior.
Also from a nonequilibrium perspective there is a broad range of different processes which can be studied with the Metropolis algorithm and which recently have attracted significant attention, such as field-driven hysteresis, phase ordering, Kibble-Zurek mechanism, Mpemba effect, etc.
From a computational point of view, these physical aspects are usually studied under the restriction to short-range interacting systems since the conventional simulation of a long-range interacting system with $N$ constituents comes along with an $O(N^2)$ complexity.
This fact has severely impeded simulation studies and only in very recent times algorithmic advances have alleviated this problem, leading to a new wave of important insights into the nonequilibrium physics of long-range interacting systems~\cite{Christiansen2019a,Christiansen2020,christiansen2020zero,Mueller2022,Mueller2024}.
One keystone for this development has been our recent fast and hierarchical Metropolis algorithm~\cite{Mueller2023} which, however, comes with a fairly elevated technical overhead constituting a barrier to its broad adoption.

In this Letter we propose a new, more streamlined algorithm.
Opposed to the fast, hierarchical algorithm~\cite{Mueller2023}, it is even applicable for systems with random signs in the interaction.
Regarding the technical requirements for its implementation, the new algorithm is on par with the conventional Metropolis algorithm, while still providing a significant reduction of the computational complexity.
Since the algorithm, using the same random-number sequence, reproduces exactly the same Markov chain as the Metropolis algorithm with the exact evaluation of the energy, all existing physical knowledge about the systems under study can be carried over one-to-one.
While this property makes it particularly useful for the investigation of nonequilibrium processes, the algorithm can, of course, also be applied to equilibrium systems.
The property of reproducing the Metropolis criterion, the simplicity of its implementation, and the coincident enormous reduction of the computational burden render it a very promising candidate for broad application in studies of a vast spectrum of phenomena enrooted in long-range interacting systems.

As prototypical models we consider three different variants of long-range $\mathrm{O}(n)$ spin models which all have in common the Hamiltonian
\begin{equation}
  \label{eq:on_hamiltonian}
 {\cal H}=-\frac12 \sum_{i}\sum_{j \neq i} J_{i,j}\mathbf{s}_i\mathbf{s}_j,
\end{equation}
where the spins $\mathbf{s}_i$ are $n$-dimensional unit vectors, including the Ising model for $n=1$.
The interactions are given by
\begin{equation}
  \label{eq:interaction}
  \left| J_{i,j}\right| = r(i,j)^{-d-\sigma},
\end{equation}
where $r(i,j)$ denotes the Euclidean distance between $\mathbf{s}_i$ and $\mathbf{s}_j$, $d$ is the spatial dimension, and $\sigma>0$ is a tuneable parameter determining the decay of the interactions.
The absolute value $\left| J_{i,j}\right|$ is taken to account for the possibility of antiferromagnetic interactions which, for instance, appear in the long-range Edwards-Anderson spin-glass model which we consider as one example.
As in our Ref.~\cite{Mueller2023} we implement periodic boundary conditions using Ewald summation~\cite{Ewald1921}, resulting in a modified set of couplings which is determined once at the beginning of the simulation and then used together with the simple minimum-image convention.
\par
To cover a broad range of features of the model class defined by Eq.~(\ref{eq:on_hamiltonian}), we consider the following cases: i) the paradigmatic ferromagnetic long-range Ising model (LRIM), ii) the long-range ferromagnetic XY model (LRXYM) being the most simple vector spin model, and iii) the long-range Edwards-Anderson spin glass (LREAM), as prototypical model where the sign of the interaction is a quenched random variable determined individually for each pair of spins $\mathbf{s}_i$ and $\mathbf{s}_j$ at the beginning of the simulation.
The former two systems fulfill translation invariance such that a single set of couplings $J_{0,k}$ can be used for the simulation.
The latter one additionally needs a matrix $\iota_{i,j}$ containing the sign of the interaction for each pair of spins.
\par
In its traditional implementation, the Metropolis algorithm is based on a proposal-acceptance scheme.
For the system which is currently in a microstate with energy $E^\mathrm{old}$ a new microstate with energy $E^\mathrm{new}$ is proposed.
The new microstate is accepted if
\begin{equation}
  \rho \leq \exp(-\beta \Delta E),
  \label{eq:standard_criterion}
\end{equation}
where $\rho \in [0,1)$ is a random number, $\beta$ the inverse temperature, and $\Delta E = E^\mathrm{new} - E^\mathrm{old}$ is the energy change associated with the proposed update.
Usually, for spin systems an update only concerns a single spin and hence the calculation of $\Delta E$ in a model with only nearest-neighbor interactions comprises a constant, very limited number of terms.
For a long-range $\mathrm{O}(n)$ model the {calculation of $\Delta E$ resulting from the update of a single spin $\mathbf{s}_i$ reads
\begin{equation}
    \Delta E = E^\mathrm{new}_i - E^\mathrm{old}_i = -\Delta \mathbf{s}_i  \sum_{j \neq i} J_{i,j}  \mathbf{s}_j,
    \label{eq:energy_change}
\end{equation}
where $\Delta \mathbf{s}_i = \mathbf{s}_i^\mathrm{new} - \mathbf{s}_i^\mathrm{old}$ and the summation runs over all other spins in the system.
Thus, it requires the summation over the whole system, resulting in the infamous $O(N^2)$ complexity per Monte Carlo sweep (MCS) consisting of a number of updates equal to the number of spins contained in the system.
However, as shown in Refs.~\cite{Schnabel2020,Mueller2023} this problem can in many cases be circumvented.
The full evaluation of (\ref{eq:energy_change}) is avoided by solving (\ref{eq:standard_criterion}) for $\Delta E$
\begin{equation}
  \Delta E \leq  -\frac{\ln{\rho}}{\beta} \equiv E^\mathrm{th},
  \label{eq:threshold_energy}
\end{equation}
defining the threshold energy $E^\mathrm{th}$ up to which an update gets accepted.
This allows decisions based upon strict bounds $\Delta E^\mathrm{min} \leq \Delta E$ and $\Delta E^\mathrm{max} \geq \Delta E$.
If $\Delta E^\mathrm{min}$ is larger than $E^\mathrm{th}$ the update is rejected, while it is accepted if $\Delta E^\mathrm{max}$ is smaller than $E^\mathrm{th}$.
Depending on the circumstances, the decisions can be taken with an extremely reduced computational effort.
In Ref.~\cite{Mueller2023} we built upon this idea, establishing a generic algorithm for both lattice and off-lattice systems.
This was achieved by leveraging the tree-structure of a hierarchical spatial decomposition of the simulation domain.
Here, we propose a very simplified framework based on partial sums combined with a predecision scheme.
To achieve this, we first introduce a permutation $\pi(j)$ in (\ref{eq:energy_change}) which allows to rearrange the summation order to best fit our algorithmic purposes (which in most cases is given by sorting $J_{i,j}$ in decreasing order with respect to their magnitude).
Secondly we split the sum into two parts,
\begin{align}
  \Delta E &= \underbrace{ -\Delta \mathbf{s}_i \sum_{j = 1}^{n} J_{i,\pi(j)}  \mathbf{s}_{\pi(j)}}_{K\equiv\,\text{known part}}  \underbrace{- \Delta \mathbf{s}_i \sum_{j=n+1}^{N-1} J_{i,\pi(j)}  \mathbf{s}_{\pi(j)} }_{R\equiv\,\text{rest}}\nonumber\\
  &= K(n) + R(n)
  \label{eq:partitioned_energy_change}
\end{align}
of which only the first sum $K(n)$ will be treated explicitly and thus is known exactly.
For the treatment of $R(n)$ we introduce an additional quantity
\begin{equation}
  U(n)
  = \left| \Delta \mathbf{s}_i \right| \sum_{j=n+1}^{N-1} \left| J_{i,\pi(j)} \right|
  = \left| \Delta \mathbf{s}_i \right| \left[ J_i^\mathrm{int} -  \sum_{j=1}^{n} \left| J_{i,\pi(j)} \right|\right],
  \label{eq:unknown-part}
\end{equation}
where
\begin{equation}
  J_i^\mathrm{int} = \sum_{j=1}^{N-1}\left| J_{i,j} \right|,
  \label{eq:j_int}
\end{equation}
is the sum over the couplings between $\mathbf{s}_i$ and all other spins in the system.
Since $J_i^\mathrm{int}$ are calculated once at the beginning of the simulation (in fact for systems with periodic boundary conditions and translation invariance there is only a single $J_0^\mathrm{int}$ which in the following will be simply denoted as $J^\mathrm{int}$),
$U(n)$ can be calculated alongside with $K(n)$ during the summation process.
We note that for each summand in $R(n)$, $-\left| \Delta \mathbf{s}_i J_{i,k} \right| \leq -\Delta \mathbf{s_i} J_{i,k}\mathbf{s}_k \leq \left| \Delta \mathbf{s}_i J_{i,k} \right|$ (because the spins are unit vectors) and thus
\begin{equation}
  \label{eq:R-bounds}
  -U(n) \leq R(n) \leq U(n).
\end{equation}
Now, we can easily establish strict lower and upper bounds on (\ref{eq:partitioned_energy_change})
\begin{align}
  \Delta E^\mathrm{min}(n) &= K(n) - U(n),\quad\text{and} \nonumber\\
  \Delta E^\mathrm{max}(n) &= K(n) + U(n),
  \label{eq:bounds}
\end{align}
which allow a predecision by prematurely terminating the sum at $n_0$ summation steps if either $\Delta E^\mathrm{min}(n_0)>E_\mathrm{th}$ (rejection) or $\Delta E^\mathrm{max}(n_0)<E_\mathrm{th}$ (acceptance).

Finally, it remains to choose an efficient permutation $\pi(j)$ such that $U(n)$ in Eq.~(\ref{eq:unknown-part}) diminishes as efficiently as possible.
The most intuitive approach, which we adopt here, is to pick a permutation that arranges $\left|J_{i,j}\right|$ in monotonously decreasing order to minimize $U(n)$ for given $n$~\footnote{An imperfect ordering in $\left|J_{i,j}\right|$, where subsequences are additionally ordered to minimize cache misses, can lead to further improved wall-clock times.}.
\par
The explicit algorithmic formulation for a system with translation symmetry contains the following preparatory steps
\begin{enumerate}
    \item Calculate all interactions $J_{0,k}$ from the reference spin at $\mathbf{r}=\mathbf{0}$ to all other spins in the lattice.
    \item Store these interactions alongside their corresponding relative position vectors $\mathbf{r}_k$.\label{it:sortieren}
    \item Sort this list in descending order for the interaction strength $J_{0,k}$.
    \item Compute the integrated interaction $J^{\text{int}} = \sum_{j=1}^{N-1} J_{0,j}$.
\end{enumerate}
Since these steps are only performed once at the beginning of the simulation, they do not significantly contribute to the algorithm's execution time.
However, the spins $\mathbf{s}_j$ now must be properly identified using $\mathbf{r}_j = \mathbf{r}_i + \mathbf{r}_k$ taking into account the minimum-image convention.
Each update then consists of the following steps:
\begin{enumerate}
\item Randomly select a spin $\mathbf{s}_i$ to update.
\item Propose a new state $\mathbf{s}_i^\mathrm{new}$ (for Ising spins $s_i^\mathrm{new} = -s_i^\mathrm{old}$)
\item Randomly draw the threshold energy $E^\mathrm{th}$ (\ref{eq:threshold_energy}).
\item If $E^\mathrm{th} > \left|\Delta\mathbf{s}_i\right|J_i^\mathrm{int}$ accept the update directly.
\item Continue summation until
  \begin{itemize}
  \item Either $\Delta E^{\mathrm{max}}(n) < E^{\mathrm{th}}$: Accept update.
  \item Or $\Delta E^{\mathrm{min}}(n) > E^{\mathrm{th}}$: Reject update.
  \end{itemize}
\end{enumerate}
\begin{figure}
  \centering
  \includegraphics{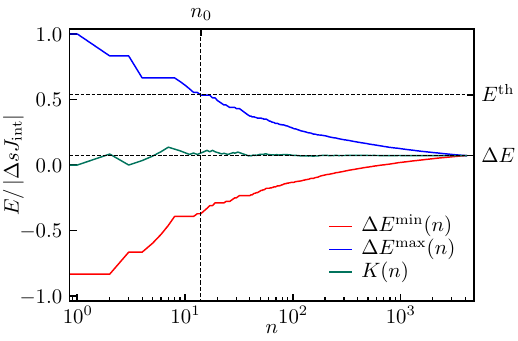}
  \caption{Exemplary illustration of the predecision process for the long-range Ising model in $d=2$ with $\sigma = 0.6$, $L=64$, and $T=20$.
    Due to the elevated temperature which is well above the critical temperature $T_c$ the system is in a disordered state.
    The green line represents the exactly known part $K(n)$ which accounts for the actual sign of the individual interactions.
    The red and the blue line show the evolution of the upper and lower bound of $\Delta E^\mathrm{min/max}(n)=K(n)\mp U(n)$, respectively which by construction converge monotonously to the exact value of $\Delta E$ associated with the update.
    Already at $n_0=14$ (of $N - 1=64^2 - 1= 4095$) summation steps the upper bound $\Delta E^\mathrm{max}$ lies below the energy threshold $E^\mathrm{th}$ such that the update can be accepted with the explicit knowledge of only a tiny fraction of all interactions.}
  \label{fig:example-decision}
\end{figure}

In Fig.~\ref{fig:example-decision} we illustrate this procedure by visualizing an exemplary decision process.
There, we plot $K(n)$ together with both $\Delta E^\mathrm{min}(n)$ and $\Delta E^\mathrm{max}(n)$.
One clearly sees how the energy bounds approach the true value of $\Delta E$ for growing index $n$.
In this specific situation, the update can be accepted already after $n_0=14$ ($\approx 0.34\%$ of total number of spins) summation steps because the difference between the threshold energy $E^\mathrm{th}$ and the actual value $\Delta E$ associated with the spin flip is fairly large.
\par
Next, we turn to a performance analysis.
The common approach to assess the computational complexity via the scaling of wall-clock times of the algorithm has the major drawback that it can strongly depend on several environmental factors.
These are primarily the hardware architecture and configuration of the compute node, and the workload that is running in parallel, which can influence boost clocks and cache availability.
While one can try to exactly reproduce the environment in which the code runs, it will still be difficult to safely assess runtimes with an uncertainty of less than 5\%.
Hence, we here choose a different approach by measuring the average number of needed summation steps $\langle n_0 \rangle$ per update, yielding an architecture and environment independent benchmark.
\par
The resulting scaling analyses are presented in Fig.~\ref{fig:scaling-plot}.
The two-dimensional LRIM and LRXYM are benchmarked at the respective transition temperatures, while for the two-dimensional LREAM we perform the simulations at $0.5 T_c(\sigma)$ of the LRIM.
We have checked, that the choice of the simulation temperature does not influence the asymptotic algorithmic scaling, rendering the choice of the simulation temperature secondary for our analysis.
\begin{figure*}
  \centering
  \includegraphics[width=\textwidth]{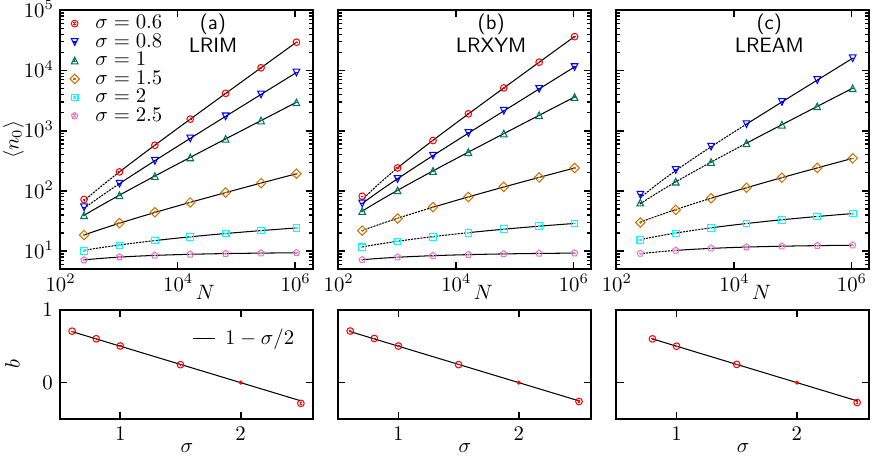}
  \caption{Scaling analysis of the average number of needed summation steps $\langle n_0 \rangle$ for the (a) LRIM, (b) LRXYM, and (c) LREAM in $d=2$ dimensions.
    The solid lines represent the fits to $\langle n_0 \rangle (N) = C + AN ^b$ in the fitting interval to extract the asymptotic algorithmic scaling and the dotted lines their continuations to smaller $N$.
    The lower panel shows the resulting exponent $b$.
    For $\sigma=2$ where we expect $b=0$ a logarithmic ansatz is employed.
    For $\sigma > 2$ the computational effort per update becomes constant for $N \rightarrow \infty$.}
  \label{fig:scaling-plot}
\end{figure*}
For all three models (LRIM, LRXYM, and LREAM) a power-law dependence of the form
\begin{equation}
  \label{eq:ansatz}
  \langle n_0 \rangle (N) = C + AN ^b,
\end{equation}
can be fitted to the data, for $\sigma \neq 2$, yielding exponents
\begin{equation}
  \label{eq:sigma-b}
  b = 1 - \sigma/2.
\end{equation}
By going into more details of the algorithm's performance and considering the probability density $f(n_0)$ and its first moment $\langle n_0 \rangle = \sum_{n_0=0}^{N-1} n_0 f(n_0)$, we have theoretical arguments~\footnote{F. Müller and W. Janke, Analysis of runtime distributions in predecision Metropolis Monte Carlo, in preparation.} that the 2 in (\ref{eq:sigma-b}) should more generally be replaced by the dimension $d$.
This is demonstrated in the Appendix for the example of $d=1$.
Hence, for $\sigma=d$ we expect a logarithmic divergence being supported by the good description of the data with the logarithmic ansatz $\langle n_0\rangle(N) = a \ln N + C$ employed for $\sigma=2$ in Fig.~\ref{fig:scaling-plot}(a-c) for all three considered models.
For $\sigma > 2$, $\langle n_0 \rangle$ approaches a constant for large $N$, i.e., in this short-range-like regime the exponent $b$ plays the role of a subleading correction.
The fit intervals are chosen individually for each fit to yield a $\chi^2$ per degree of freedom close to unity.
The resulting fits for all models and values of $\sigma$ in the chosen fit interval are shown as solid black lines while the dotted lines represent their continuation to smaller $N$ which were excluded from the fit.
We thus find that the computational complexity is largely model independent, while the prefactors are model-dependent, varying by factors $<2$.
For $\sigma\geq 1.5$, compared to a simulation with a full calculation of the energy differences, already for $L=1024$ we observe a reduction of the needed summation steps by factors between 3000 and 50\,000 which due to the better scaling would further increase with growing system size.
Hence, the algorithm enables simulations within days which would take decades with a naive implementation.
\par
To conclude, we have presented a new algorithm for the exact treatment of long-range interaction for Metropolis Monte Carlo simulation of lattice systems.
The algorithm can be thought of as a simulation with an adaptive cutoff which for each individual update is chosen just large enough to take the correct decision according to the Metropolis criterion.
In a model-independent fashion, we observe a reduction of the computational complexity from $O(N^2)$ to $O(N^{2-\sigma/d})$ for $0 < \sigma < d$ and to $O(N)$ for $\sigma > d$, implying that for $\sigma > d$ even asymptotically it outperforms the technically significantly more challenging algorithm presented in Ref.~\cite{Mueller2023}.
For smaller values of $\sigma$ close to $d$ it can still be competitive up to very large system sizes, due to the algorithmic scaling remaining weak and the algorithmic prefactor being significantly smaller due to the reduced complexity of the algorithm.
\par
We have not investigated any dynamical properties since the decisions of the Metropolis algorithm are exactly reproduced and thus also its dynamical properties.
The new method is very easy to implement and can easily be adapted to other use-cases.
Finally, we also provide a working implementation in the Supplementary Material~\footnote{See Supplemental Material at \{will be added\} where a full implementation of the algorithm in the Julia programming language is provided.}, such that the algorithm can easily be adopted and extended by other researchers in the field.
Possible applications contain equilibrium studies of frustrated systems for which no efficient cluster algorithm is available such as, e.g., spin glasses, random-field models, or the dipolar Ising model~\cite{DeBell2000}.
Additionally, its features render the algorithm well suited for the investigation of nonequilibrium processes such as, e.g., phase-ordering kinetics~\cite{Bray1994}, the Kibble-Zurek mechanism~\cite{Du2023} or the Mpemba effect~\cite{Teza2025}, and many others~\cite{Odor2004}.

\begin{acknowledgments}
  This project was supported by the Deutsch-Französische Hochschule (DFH-UFA) through the Doctoral College ``$\mathbb{L}^4$'' under Grant No.\ CDFA-02-07.
\end{acknowledgments}

\appendix
\section{End Matter: Algorithmic scaling in $d=1$}
In the main text we have stated that the denominator 2 in (\ref{eq:sigma-b}) should actually read $d$.
To corroborate on this statement we additionally perform a scaling analysis of $\langle n_0 \rangle$ for the LRIM in $d=1$.
As for the $d=2$ LRIM the simulations are again performed at $T_c(\sigma)$.
Since for $\sigma>1$ the transition temperature is zero, the system in equilibrium is always in the ground state and the simulations become trivial.
Therefore, we here restrict $\sigma$ to $\sigma\leq 1$.
As demonstrated in Fig.~\ref{fig:scaling1D}, the exponents in the scaling ansatz~(\ref{eq:ansatz}) follow the relation $b=1-\sigma$ confirming that the general expression for $b$ should read $b=1-\sigma/d$.
\begin{figure}
  \centering
  \includegraphics{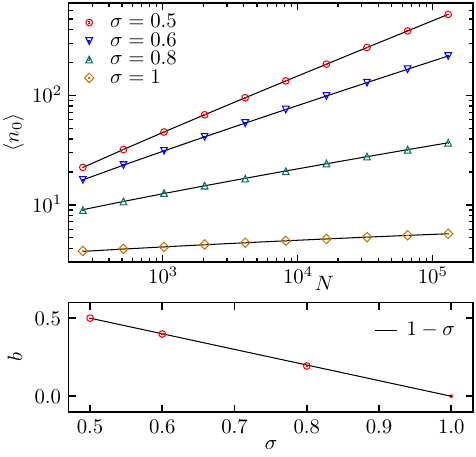}
  \caption{Scaling analysis of the average number of needed summation steps $\langle n_0 \rangle$ for the $d=1$ LRIM.
    The solid lines represent the fits to $\langle n_0 \rangle (N) = C + AN ^b$ which in this case are performed over the full range of system sizes.
    The lower panel shows the exponent $b$ of the power-law ansatz describing the asymptotic growth of $\langle n_0 \rangle$.
    For $\sigma=1=d$ where we expect $b=0$ a logarithmic ansatz is taken.}
  \label{fig:scaling1D}
\end{figure}
\end{document}